# Dynamic Partial Computation Offloading for the Metaverse in In-Network Computing

Ibrahim Aliyu[1], Member, IEEE, Seungmin Oh[1], Namseok Ko[2], Tai-Won Um[3], and Jinsul Kim[1], Member, IEEE
[1]Department of ICT Convergence System Engineering, Chonnam National University, Gwangju 61186, Korea.
[2]Telecommunication Media Laboratory, Electronics and Telecommunications Research Institute, Daejeon 34129, Korea.
[3]Graduate School of Data Science, Chonnam National University, Gwangju 61186, Korea

Corresponding authors: Tai-Won Um (email: stwum@jnu.ac.kr) and Jinsul Kim (email: jsworld@jnu.ac.kr).

This work was partly supported by the Electronics and Telecommunications Research Institute (ETRI) grant funded by the Korean government [23ZR1110, "A study of hyper-connected thinking Internet technology by autonomous connecting, controlling and evolving ways"]; the Institute of Information and Communications Technology Planning and Evaluation (IITP) grant funded by the Korean government (MSIT) [2021-0-02068, Artificial Intelligence Innovation Hub]; and the National Research Foundation of Korea (NRF) grant funded by the Korean government (MSIT) [NRF-2021R1I1A3060565].

**ABSTRACT** The computing in the network (COIN) paradigm is a promising solution that leverages unused network resources to perform tasks to meet computation-demanding applications, such as the metaverse. In this vein, we consider the partial computation offloading problem in the metaverse for multiple subtasks in a COIN environment to minimize energy consumption and delay while dynamically adjusting the offloading policy based on the changing computational resource status. The problem is NP-hard, and we transform it into two subproblems: the task-splitting problem (TSP) on the user side and the task-offloading problem (TOP) on the COIN side. We model the TSP as an ordinal potential game and propose a decentralized algorithm to obtain its Nash equilibrium (NE). Then, we model the TOP as a Markov decision process and propose the double deep Q-network (DDQN) to solve for the optimal offloading policy. Unlike the conventional DDQN algorithm, where intelligent agents sample offloading decisions randomly within a certain probability, the COIN agent explores the NE of the TSP and the deep neural network. Finally, the simulation results reveal that the proposed model approach allows the COIN agent to update its policies and make more informed decisions, leading to improved performance over time compared to the traditional baseline.

## I. INTRODUCTION

The metaverse is a persistent and immersive simulated environment breaching the physical and digital world through virtual, augmented, and extended reality (XR) [1]. An essential characteristic of the metaverse is its potential to provide an immersive experience to large groups of users who simultaneously share a keen sense of mutual presence through XR. With the proliferation of cheap XR equipment and technology, such as head-mounted displays, the metaverse is expected to feature in multiplayer games, workplace meetings, scientific research, and engineering [2].

However, deploying portable and small devices, such as small head-mounted displays, mobile augmented reality, and others that can meet the demands of XR, is still a significant challenge because such devices are limited in processing power, storage, and battery life [3, 4]. Although mobile edge computing (MEC) offers a solution through remote task offloading (TO), it cannot simultaneously support massive user demand [5-8]. The computing in the network (COIN) paradigm is a promising solution that leverages unused network resources to perform tasks, reducing delay and satisfying the quality of experience (QoE) requirements [9-11]. However, adding computing resources or enabling COIN increases power consumption. Allocating the dynamic COIN computing resources and meeting constantly changing user demands alongside the availability is a critical problem. Considering that XR functionality can be split into several subtasks [3], the task-splitting problem (TSP) from the user perspective and subsequent optimization of dynamic COIN resource allocation under delay and energy constraints is a fundamental problem.

## A. SCENARIO

We consider a metaverse scenario where multiple users with user equipment (UE) constraints participating in an XR social event, such as shopping, competing for the available computing resources. Due to the UE(s) entering and leaving the network and the dynamic demand and changing status of the resources, an optimal offloading policy must be predicted and updated accordingly. We examined the XR functionalities that can be split between the UE and COIN resources. The COIN resources are expected to spread across the nodes; thus, we classify the resources into two: fog in-network computing nodes (FINs) and edge in-network computing nodes (EINs). The FINs are closer to the UE, whereas the EINs are farther from the user. These nodes are controlled and managed by a COIN controller, which knows the status of the nodes, including the request queue and distances. The spread of the computing nodes offers the benefit of distributed computing. However, it creates the challenge of determining an optimal offloading location considering the dynamic user demand and requests pending in the network and each user's task-splitting (TS) decision. Thus, we model the TSP as an ordinal potential game (OPG) and obtain its NE. Each user makes an independent TS decision based on the generated task; thus, the solution is suboptimal. Thus, we employ the double deep Q-network (DDQN) on the network side to optimize the offloading policy over time.

## B. MOTIVATION AND CONTRIBUTIONS

The massive deployment of the metaverse is expected to generate tremendous demand for computing resources because it could transform how we live, work, and interact with the physical world [12]. Moreover, UE, such as mobile devices, is resource-constrained to provide delay-sensitive and compute-intensive services for the metaverse [5]. Conventional tasks can be fully or partially offloaded remotely to MEC at the network edge or a cloud data center to address this challenge. However, MEC is often limited and cannot support the massive demand of simultaneous users [5-8]. Because the metaverse is expected to provide an immersion experience to a large group of simultaneous users [13], optimal computing resource allocation is critical to resolving the resource demand conflict among simultaneous users. The COIN paradigm is a promising solution that leverages unused network resources to perform tasks, reducing delay and meeting QoE requirements [9, 10]. However, adding computing resources or enabling COIN increases power consumption. This competing situation leads to a joint optimization problem of time delay and energy consumption for metaverse tasks. Although tremendous progress has been made in solving the network joint task-offloading problem (TOP) [14-16], most studies have focused on atomic TO. Furthermore, more research is needed to understand various metaverse tasks and offloading to COIN-enabled nodes with the corresponding resources required for each task.

Therefore, this paper addresses the problem of metaverse applications, where resource constraints and the demand for real-time computation present a significant challenge. We focus on metaverse multiple subtask offloading and the subsequent offloading policy optimization, considering the dynamic demand and nature of the COIN resources under energy and delay constraints. This study considers the metaverse TSP, partial TOP, and computing and caching resource allocation for massive deployment. Specifically, we formulate a joint metaverse task for partial computational offloading in a COIN environment to minimize energy consumption and delay while dynamically adjusting the offloading policy based on changing the demand for computational resources to avoid congestion.

We prove that the problem is NP-hard, so we transform it into two subproblems: an optimal TSP on the user side and a computational TOP on the COIN side. We model the TSP as an OPG and propose a decentralized algorithm to obtain its Nash equilibrium (NE). The users in the OPG act independently and not toward the collective goals of the system; thus, its solution is suboptimal for online TS decisions for the metaverse.

Hence, we formulate the suboptimal performance resulting from each user in the OPG as a Markov decision process (MDP) to solve for the optimal partial TO policy. We propose a DDQN to optimize the system. The agent learns to adjust the uploading policy to meet the global demand and resources available in the network. Unlike the conventional DDQN algorithm, where intelligent agents sample offloading decisions randomly within a certain probability, the COIN agent explores the NE of the TS algorithm and the deep neural network (DNN) within the probability. By exploring the TS game solution, the agent can update its policies to make more informed decisions in the future, leading to improved performance over time.

The proposed approach performs better in lowering partial TO costs and optimizes the future offloading policy against the baseline approaches. The main contributions of this paper are as follows:

1) We formulate a joint metaverse TS and computational offloading in a COIN environment to minimize energy consumption and delay while dynamically adjusting the offloading policy based on the demand to avoid congestion. We prove that the problem is NP-hard.
2) We decompose the problem into two parts to solve the problem: TSP on the user side and computational TOP on the COIN side. Considering that the TSP is a combinatorial optimization over a multidimensional discrete space, we formulate the problem as an OPG and prove the existence of the NE to determine the optimal TS for each user.

3) To address the suboptimal performance resulting from each user acting noncooperatively to minimize their own cost, not the system state's collective goal, we model the TOP as an MDP. We propose a DDQN to optimize the system. The agent learns to adjust the uploading of the proposed DDQN to learn the optimal offloading policy under the users' TS game solution to address the TOP. The agent learns to adjust the uploading policy to meet the global demand and resources available in the network.
4) We conduct extensive simulations to evaluate the dynamic TO and resource allocation performance.

The rest of the paper is organized as follows. Section II discusses related studies, and Section III presents the system model, including the problem formulation for the TSP and TOP. Next, Section IV discusses a dynamic TO scheme to solve the problem. Then, Section V discusses the simulation results to verify the effectiveness of the proposed method. Section VI concludes the paper.

## II. RELATED WORK

The TO can be classified as full or partial (partitioning) offloading [17]. Full offloading focuses on atomic tasks. Partial offloading deals with divisible tasks or applications. Partial offloading can be further classified as data-oriented, continuous-execution, and code-oriented partitioning offloading [18]. In data-oriented partitioning offloading, the application can split the task into subtasks with the number of data known beforehand [19]. In contrast, in continuous-execution partitioning offloading, the number of data is unknown [20]. Code-oriented partitioning offloading focuses on applications where tasks can be divided into subtasks with dependencies [21, 22]. Ding, et al. [23] considered code-oriented partitioning offloading in multiuser and multi-MEC scenarios with the aim of minimizing the execution overhead.

Joint communication and computing resource allocation problems for TO have recently received considerable attention, particularly in MEC networks [14-16]. For instance, Jošilo and Dán [24] considered the problem of offloading latency-sensitive computation tasks in edge computing under network slicing. Inter- and intraslice radio and computing resource management have been investigated for low-complexity dynamic resource allocation. In addition, Hu, et al. [25] addressed atomic TO optimization in MEC, where multiple users compete for resources while minimizing transmission power using greedy-pruning algorithms. Tong, et al. [26] proposed the Lyapunov online energy consumption optimization algorithm to solve the queue backlog and energy consumption in MEC. Further, Liu, et al. [27] addressed TO in MEC considering the mobility user devices and proposed a mobility-aware and code-oriented partitioning computational offloading scheme.

Moreover, other research has considered offloading tasks to the three major resources—caching, communication, and computing resources [16, 28, 29]. However, MEC is often limited and cannot support the massive demand of simultaneous users [4]. This problem warrants a solution that can reduce the computational burden of MEC.

In this vein, the COIN paradigm has also garnered interest as a potential solution to support latency-constrained applications. Cooke and Fahmy [30] demonstrated the significant efficiency of COIN for distributed streaming applications in terms of latency, throughput, bandwidth, energy, and cost. The use of machine learning, such as the decision tree, multilayer perceptron, and support vector machine, to orchestrate delay-constrained task placement in COIN has also displayed remarkable performance [31].

Furthermore, COIN-enabled holographic streaming applications present improved performance in terms of low latency and high bandwidth [11, 32]. These studies have demonstrated the potential of COIN to support the massive deployment of latency-contained applications, such as the metaverse. However, for optimal performance and load distribution in terms of energy and latency, considering the location of the COIN node and user demand under changing network conditions, a dynamic TO scheme is essential.

Despite this, most previous studies [6], [31], [33-37] have treated tasks as a single unit and did not consider situations where the tasks could be divided and handled by different computing nodes. This situation is critical in the metaverse, where a metaverse task consists of multiple tasks that can be decomposed and offloaded to different computing nodes (e.g., a COIN node).

Although one study [3] proposed XR TS in the fifth-generation (5G) network, their solution offers three upload modes. This approach may be sufficient for a virtual, augmented, or XR experience with few users, but for a massive metaverse deployment scenario, optimizing the TO mode is essential, considering the tremendous simultaneous demand for network resources.

Similarly, Tütüncüoğlu, et al. [38] considered subtask offloading in server-less edge computing and proposed an online learning algorithm maximizing the application utility. In addition, Zhang, et al. [39] addressed the subtask dependency offloading problem in MEC and proposed a scheme that minimizes the subtask energy and latency execution. Nonetheless, these studies considered binary offloading decision-local or server-less edge computing or MEC and a static topology with users always accessing the same resources.

Motivated by the limitations of the mentioned work, this study addresses the problem of metaverse applications, where resource constraints and the demand for real-time computations considering multiple subtasks present significant challenges. Moreover, optimal resource allocation to efficiently harness the COIN paradigm, considering its dynamic resources, is a unique problem. Thus, we present a novel approach that considers both metaverse TS on the user side and computational TO on the COIN side, aiming to minimize energy consumption and delay. We propose a decentralized algorithm to obtain the NE of the TSP, which is an interesting contribution that addresses the challenge of users acting noncooperatively. Contrary to previous studies, the DDQN algorithm explores the NE of the TS algorithm, leading the agent to make more informed decisions in the future and improving performance over time.

TABLE I
RELATED WORK

| Related work | Method | Task splitting | Delay | Energy | Cache | Metaverse | COIN |
|---|---|---|---|---|---|---|---|
| 2019 [23] | Convex optimization | ✓ | ✓ | ✓ | ✗ | ✓ | ✗ |
| 2020 [25] | Game theory | ✗ | ✓ | ✓ | ✗ | ✗ | ✗ |
| 2020 [30] | Hardware, software | ✓ | ✓ | ✓ | ✗ | ✗ | ✓ |
| 2020 [33] | RL and game theory | ✗ | ✓ | ✓ | ✗ | ✗ | ✓ |
| 2020 [34] | Convex optimization | ✗ | ✓ | ✓ | ✗ | ✗ | ✗ |
| 2021 [40] | Game theory | ✓ | ✗ | ✗ | ✓ | ✗ | ✗ |
| 2021 [41] | Berger model, service justice | ✓ | ✗ | ✗ | ✓ | ✗ | ✗ |
| 2022 [27] | Convex optimization and Lagrangian approach | ✓ | ✓ | ✓ | ✗ | ✓ | ✗ |
| 2022 [26] | Lyapunov optimization | ✗ | ✗ | ✓ | ✗ | ✗ | ✗ |
| 2022 [6] | Game theory, RL | ✗ | ✗ | ✓ | ✓ | ✗ | ✗ |
| 2022 [31] | DT, MLP, SVM | ✗ | ✓ | ✗ | ✗ | ✗ | ✓ |
| 2022 [32] 2023 [11] | Network slicing | ✗ | ✓ | ✗ | ✗ | ✗ | ✓ |
| 2023 [36] | Game theory | ✗ | ✓ | ✓ | ✗ | ✗ | ✗ |
| 2023 [37] | RL | ✗ | ✓ | ✓ | ✓ | ✓ | ✗ |
| **Our proposal** | **Game theory, RL** | ✓ | ✓ | ✓ | ✓ | ✓ | ✓ |

Table 1 lists the comparisons between the proposed method and state-of-the-art schemes involving partial computational offloading. As evident, most previous work has focused on atomic tasks in MEC. However, the metaverse is characterized by multiple subtasks whose computational offloading is critical. Moreover, the dynamic set of COIN resources and changing user demand require attention to harness the benefit of the COIN paradigm. Although one study [6] employed the game theory and reinforcement learning (RL) approach to address software caching updates and computational offloading, the scope is only for MEC. The TO is handled on the user side using the game theory, whereas the RL operates at the MEC end to predict whether the task software is already cached in the server. No further optimization of the offloading policy is conducted to handle dynamic user demand and the changing resource status. The uniqueness of this approach lies in the decoupling of the TSP on the user side and the subsequent TO policy optimization.

## III. SYSTEM MODEL

### A. NETWORK MODEL

This study considers XR applications to be a metaverse constituent with growing interest across a spectrum of users. The XR processing entails eight functionalities, which can be grouped into four components: object tracking and detection, simultaneous localization and mapping (SLAM) and map optimization with a point cloud dataset, hand gesture and pose estimation, and multimedia processing and transport, such as rendering and encoding [3] (see Fig. 1). The critical notation used in this article is summarized in Table 2.

In this scenario, a group of users with a metaverse application simultaneously generates fork-join-type job tasks on each device. Each task can be split into four subtasks (four major components) that can be executed in parallel, in series, or combined. The UE can locally perform the task or offload it to an FIN or EIN considering the constraints of power, computing capacity, QoE, and demand from other users in the network, as illustrated in Fig. 2. The COIN concept guarantees computing resources to be added or made available in the network. Thus, we consider the availability of computing resources alongside the joint optimization problem.

We consider a multiuser network system model for the COIN network, which consists of a UE set $\kappa = \{1, 2, ... K\}$, a set of fifth- or fourth-generation access points $\mathcal{A} = \{1,2,..A\}$ through which the UE can offload tasks, a FIN set $FIN = \{1, ......fin\}$ and an EIN set $EIN = \{1...2,...ein\}$. The COIN (communication and computing) resources are defined as the set $\mathcal{R} = \mathcal{A} \cup FIN \cup EIN$.

| Functionalities | Major components | |
|---|---|---|
| Object tracking | Object tracking and detection | Multimedia processing and transport (rendering, syn and encoding) |
| Object detection | | |
| Map optimization | SLAM with point cloud datasets | |
| Mapping | | |
| Localization | | |
| Point cloud dataset | | |
| Sensors | Hand gesture and pose-estimation | |

**FIGURE 1.** Typical extended reality (XR) task processing on a device.

TABLE I
NOTATION SUMMARY

| Notation | Definition |
| --- | --- |
| $K, F, M$ | Number of users, number of tasks, number of subchannels |
| $\mathcal{K}, \mathcal{F}, \mathcal{M}, \mathcal{A}$ | User set, task set, subchannel set, access point set |
| $L, FIN, EIN$ | Local computing, Fog COIN, edge COIN |
| $I_f, V_f, P_f$ | Task $f$ input size, task $f$ software volume, task $f$ computation load |
| $\omega_{k,t}$ | Data transmission rate of user $k$ in slot $t$ |
| $\rho_k, F_k^L, F_k^{FIN}, F_k^{EIN}$ | User $k$ transmission power, user $k$ CPU, FIN computing capacity, EIN computing capacity |
| $\delta_{FIN}, \delta_{EIN}$ | FIN cache size, EIN cache size |
| $S_{k,t}$ | User $k$ TS decision in slot $t$ |
| $b_f^{(t)}$ | The offloading destination for task $f$ in slot $t$ |
| $\mathcal{R}_{k,t}$ | User $k$ received interference in slot |

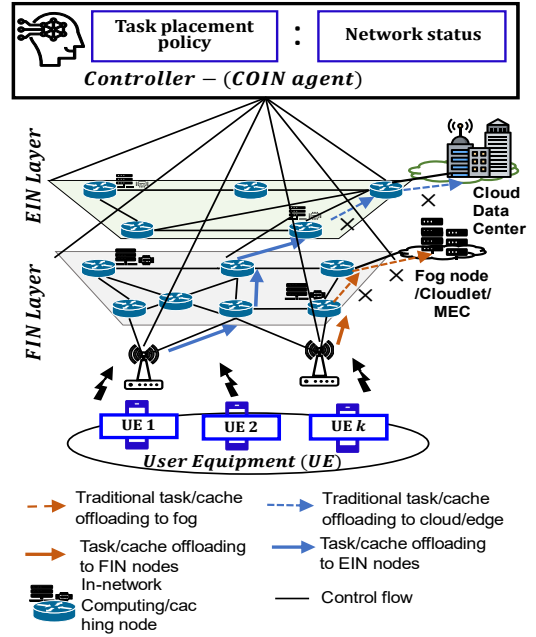

**FIGURE 2.** Metaverse task offloading in computing in the network (COIN).

### B. TASK MODEL
Unlike previous studies [6], [31],[33-36, 42] focusing on atomic tasks, metaverse divisible tasks are considered. For simplicity, we assume the subtasks are independent—a fork-joint task type. Thus, for UE $k$, we model a task of type $k$ as a directed acyclic graph $\mathcal{G}_k = (\mathcal{V}_k, E_k)$, where $v \epsilon \mathcal{V}_k$ is the subtask. The source node $v_0^k \epsilon \mathcal{V}_k$ and sink node $v_{|\mathcal{V}_k|}^k$ represent the task input transmitted via AP $\alpha \epsilon \mathcal{A}$ and the last execution subtask node, respectively. Node $v \epsilon \mathcal{V}_k \backslash \{v_0^k\}$ is the computational task and corresponds to the divisible task function constituting a metaverse task. Moreover, a directed edge $e(v_p^k, v_q^k) \epsilon E_k$ indicates that $v_p^k$ (the parent node) is executed before $v_q^k$ (the child node).

We assume each metaverse task $f \epsilon \mathcal{F}$ is denoted by the tuple parameters: $\langle I_f, V_f, P_f \rangle$, where $I_f, V_f,$ and $P_f$ are the task $f$ input size, software volume, and computational load, respectively. The task $f = \{f_{k,0}, f_{k,1}, \ldots \ldots f_{k,j}\}$, where $k$ is the user and $j$ is the number of subtasks.

### C. COMMUNICATION MODEL
This work considers that the computing nodes in the FIN and EIN operate at different frequencies and have no mutual interference between the two offloading destinations. Users offloading to the same destination experience interference because code division multiple access enables user offloading to the same destination to occupy the same spectrum of resources. Similar to another study [6], we investigated the wireless interference model for partial computation offloading in COIN. The uplink rate for the FIN and EIN, respectively, for a user $k$ in time $t$ is [6, 43]:

$$\omega_{k,t} = \frac{B}{M} \log\left(1 + \frac{\rho_k^{FIN} \eta_k^{FIN}}{\sum_{n \epsilon \mathcal{K} \backslash \{k\}, \alpha_{n,t} = \alpha_{k,t}} \rho_n \eta_n + \sigma^2}\right), \quad (1)$$

$$\omega_{k,t} = \frac{B}{M} \log\left(1 + \frac{\rho_k^{EIN} \eta_k^{EIN}}{\sum_{n \epsilon \mathcal{K} \backslash \{k\}, \alpha_{n,t} = \alpha_{k,t}} \rho_n \eta_n + \sigma^2}\right), \quad (2)$$

where $\rho_k^{FIN}$ and $\rho_k^{EIN}$ denote the transmission power for user $k$ to the FIN and EIN, $\eta_k^{FIN}$ and $\eta_k^{EIN}$ represent the channel gains for the two offloading destinations FIN and EIN, $\sigma^2$ is the variance of the complex white Gaussian channel noise, and $\sum_{n \epsilon \mathcal{K} \backslash \{k\}, \alpha_{n,t} = \alpha_{k,t}} \rho_n \eta_n$ inidicate the interference of user $k$ induced by other users. Users incur more interference using the uplink transmission rate, and a low transmission rate indicates numerous users offloading through the same channel and destination. This approach increases the energy consumption and cost of offloading tasks.

### D. COMPUTATION MODEL
We consider the local computation by the UE, FIN, and EIN. In the following, we elaborate on the three computation destinations.

#### 1) LOCAL COMPUTING
The time taken to execute a given task $f$ locally only includes the processing time on the local node at time $t$ and is defined as

$$T_{k,f}^L = \frac{P_f}{F_k^L} \quad (3)$$
$$\text{s.t } F_i^L(t) \leq F_{i,max},$$

where $F_k^L$ is the actual frequency at which a local node can execute the task.
The energy consumption of executing a given task $f$ locally at a frequency $F_k^L$ is linearly proportional to the square of $F_k^L$.

Considering the dateline constraint $\tau$ in which user $k$ must accomplish the tasks, the central processing unit (CPU) frequency for user $k$ satisfies $F_k^L \geq P_f/\tau$. Thus, the energy consumption when user $k$ executes task $f$ locally is

$$\mathcal{E}_{k,f}^L = \lambda (F_k^L)^2 P_f = \lambda \frac{P_f^3}{\tau^2}, \quad (4)$$

where $\lambda$ denotes the energy coefficient of the user device determined by the chip architecture [35, 44].

Similar to previous studies [34, 36], the cost model of user $k$ is modeled in terms of the completion time and energy consumption as follows:

$$\mathcal{C}_{k,f,t}^L = f(\delta_k^T T_k^L, \delta_k^E \mathcal{E}_k^L) = \delta_k^T T_k^L + \delta_k^E \mathcal{E}_k^L, \quad (5)$$

where $\delta_k^T$ and $\delta_k^E$ represent the delay and energy preference (weight), respectively, and $0 \leq \delta_k^T, \delta_k^E \leq 1, \delta_k^T + \delta_k^E = 1$.

2) FOG COIN

If a user decides to offload task $f$ at each time $t$, the user must offload the task input $(I_f)$ and its corresponding software $(V_f)$. Offloading a task $f$ to the external FIN consists of three delay parts. The first part deals with the time needed to transmit the input data $I_f$ and is expressed as

$$T_{k,FIN}^t = \frac{I_f + V_f}{\omega_{k,t}} \quad (6)$$

s.t $T_i^F(t) \leq \theta_{i,F}(t)$,

where $\omega_{k,t}$ is defined by (1).

The second part deals with the time taken to execute the task in the external resource FIN and is expressed as

$$T_{k,FIN}^{exe} = \frac{P_f}{F_k^{FIN}}, \quad (7)$$

where $F_k^{FIN}$ is the actual frequency at which the FIN can execute the task.

The last part of the delay is the time taken to return the computational results from the FIN to the mobile/local node. However, this delay is negligible compared to the input data $I_f$ [44]. Thus, the total delay for offloading the task to the FIN over an access point $a$ is defined as

$$T_{k,f,t}^{FIN} = T_{k,FIN}^{exe} + T_{k,FIN}^t$$
$$= \frac{P_f}{F_k^{FIN}} + \frac{I_f + V_f}{\omega_{k,t}}. \quad (8)$$

The task must be accomplished in the current time slot $t$; thus, the delay satisfies $T_{k,f,t}^{FIN} \leq \tau$. This delay is applicable during the TS problem from the user's perspective.

The TO problem aims to minimize the overall system cost; thus, additional delay at FIN is modeled using the M/M/1 queuing model. The queuing system considers the delays incurred by offloading policies of all users on the network side, including the waiting time a subtask spends in the queue before execution. The queuing delay model consists of the waiting time and delay due to utilization. The waiting time is the time a subtask waits in the queue before being served by the FIN and includes the sum of the remaining service time for all subtasks in the queue and the service time of the current subtask. The delay due to utilization considers the influence of the utilization factor on the queue delay-utilization factor of 1, indicating the FIN is fully used and the subtask experiences the full remaining service time plus its own service time as the queue delay. A utilization factor of less than 1 indicates that the FIN is not fully used. The model is described as follows.

We let $q_{FIN}$ denote the FIN queue buffer. From service time $s_{FIN}(t) = P_f/F_k^{FIN} + (I_f + V_f)/\omega_{k,t}$, the arrival time is $a_{FIN}(t) = 1/s_{FIN}(t)$. The FIN rate is given as

$$r_{FIN}(t) = \begin{cases} \frac{1}{\sum_j q_{FIN}(j) + s_{FIN}(t)}, & \text{if } q_{FIN} > 0 \\ \infty, & \text{otherwise} \end{cases}.$$

Hence, the utilization factor is

$$u_{FIN} = \begin{cases} \frac{a_{FIN}(t)}{r_{FIN}(t)}, & \text{if } r_{FIN}(t) \neq \infty \\ 1, & \text{otherwise} \end{cases}.$$

The queue delay is calculated as

$$Q_{f,t}^{FIN} = \begin{cases} \sum_{j \subseteq f} q_{FIN}(j) + s_{FIN}(t), & \text{if } u_{FIN} = 1 \\ \frac{u_{FIN}^2}{1 - u_{FIN}} s_{FIN}(t), & \text{otherwise.} \end{cases} \quad (9)$$

Thus, the total latency cost for offloading to the FIN is the summation of the processing, transmission time, and queuing delay.

$$T_{k,f,t}^{FIN} = \frac{P_f}{F_k^{FIN}} + \frac{I_f + V_f}{\omega_{k,t}} + Q_{f,t}^{FIN}, \quad (10)$$

where the first and second part on the right-hand side (RHS) of Eq. (10) is computed by Eq. (8).

The energy consumption is characterized by the energy used in offloading the task input size and software volume, considering the energy for connection scanning and execution is negligible. The energy consumption can be expressed as follows:

$$\mathcal{E}_{k,f,t}^{FIN} = \rho_k \frac{I_f + V_f}{\omega_{k,t}}, \quad (11)$$

where $\rho_k$ denotes the transmission power of the device, and $\omega_{k,t}$ is defined in (1). The total cost of offloading a task $f$ considering the delay and energy to the cost model of user $k$ is modeled in terms of the completion time and energy consumption time as follows:

$$\mathcal{C}_{k,f,t}^{FIN} = f(\delta_k^T T_{k,f,t}^{FIN}, \delta_k^E \mathcal{E}_{k,f,t}^{FIN})$$
$$= \delta_k^T T_{k,f,t}^{FIN} + \delta_k^E \mathcal{E}_{k,f,t}^{FIN}, \quad (12)$$

where $\delta_k^T$ and $\delta_k^E$ represent the delay and energy weight, respectively, as expressed in Eq. (5).

3) EDGE COIN

When the user $k$ offloads task $f$ to EIN for execution in slot $t$, it must offload both the task input $(I_f)$ and its corresponding software $(V_f)$. The total delay, which consists of execution and transmission delay, is expressed as

$$T_{k,f,t}^{EIN} = T_{k,EIN}^{exe} + T_{k,EIN}^t$$
$$= \frac{P_f}{F_k^{EIN}} + \frac{I_f + V_f}{\omega_{k,t}}, \quad (13)$$

where $F_k^{EIN}$ is the actual frequency for the EIN, and $\omega_{k,t}$ is given by Eq. (2). The first part of the RHS of (11) is the execution delay, and the second part of the RHS of (11) is the transmission delay.

Similar to Fog COIN, the additional delay at the FIN for the TO problem is modeled using the M/M/1 queuing model. The queuing delay model also consists of the waiting time and delay due to utilization and is described as follows. We let $q_{EIN}$ denote the FIN queue buffer. From service time $s_{EIN}(t) = P_f/F_k^{EIN} + (I_f + V_f)/\omega_{k,t}$, the arrival time is $a_{EIN}(t) = 1/s_{EIN}(t)$. The EIN rate is defined as

$$r_{EIN}(t) = \begin{cases} \frac{1}{\sum_j q_{EIN}(j) + s_{EIN}(t)}, & \text{if } q_{EIN} > 0 \\ \infty, & \text{otherwise} \end{cases}.$$

The utilization factor is obtained as

$$u_{EIN} = \begin{cases} \frac{a_{EIN}(t)}{r_{EIN}(t)}, & \text{if } r_{EIN}(t) \neq \infty \\ 1, & \text{otherwise} \end{cases}.$$

The queue delay is calculated as

$$Q_{f,t}^{EIN} = \begin{cases} \sum_{j \subseteq f} q_{EIN}(j) + s_{EIN}(t), & \text{if } u_{EIN} = 1 \\ \frac{u_{EIN}^2}{1 - u_{EIN}} s_{EIN}(t), & \text{otherwise.} \end{cases} \quad (14)$$

In this case, the total latency cost for offloading to the EIN is the summation of the processing, transmission time, and queuing delay:

$$T_{k,f,t}^{EIN} = \frac{P_f}{F_k^{EIN}} + \frac{I_f + V_f}{\omega_{k,t}} + Q_{f,t}^{EIN}, \quad (15)$$

where the first and second parts on the RHS of Eq. (15) are computed by Eq. (13).

Similarly, the EIN must accomplish the task in the current time slot $t$; thus, the delay transmission is modeled as follows:

$$\mathcal{E}_{k,f,t}^{EIN} = \rho_k \frac{I_f + V_f}{\omega_{k,t}}. \quad (16)$$

The total cost of offloading a task $f$ to the EIN is

$$\begin{aligned} \mathcal{C}_{k,f,t}^{EIN} &= f(\delta_k^T T_{k,f,t}^{EIN}, \delta_k^E \mathcal{E}_{k,f,t}^{EIN}) \\ &= \delta_k^T T_{k,f,t}^{EIN} + \delta_k^E \mathcal{E}_{k,f,t}^{EIN}. \end{aligned} \quad (17)$$

### E. JOINT OPTIMIZATION PROBLEM FORMULATION

We aimed to minimize the average task execution cost of all users in the network over each time slot by jointly optimizing TS decisions at the user end and TO policy on the COIN side to dynamically adjust the offloading policy to balance the TO distribution, avoiding congestion. The cost of user $k$ at the time $t$ is formulated as

$$\mathcal{C}_{k,t} = \sum_f \mathbb{1}(\mathcal{S}_{k,t} = 0) \mathcal{C}_{k,f,t}^L$$
$$+ \mathbb{1}(\mathcal{S}_{k,t} \in \mathcal{M})\left((1 - b_f^{(t)})\mathcal{C}_{k,f,t}^{FIN} + b_f^{(t)}\mathcal{C}_{k,f,t}^{EIN}\right), \quad (18)$$

where $\mathbb{1}(.)$ is an indicator function, which is 1 when the parentheses are valid; otherwise, it is 0. In Eq. (18), the first case is for the local execution ($\mathcal{S}_{k,t} = 0$) of task $f$, where the total execution cost is the local execution cost: $\mathcal{C}_{k,t} = \mathcal{C}_{k,f,t}^L$.

The second case is when user $k$ executes the task by offloading it to the FIN ($\mathcal{S}_{k,t} \in \mathcal{M}$ and $b_f^{(t)} = 0$) in which the cost, $\mathcal{C}_{k,t} = \mathcal{C}_{k,f,t}^{FIN}$, consists of the cost of transmission and execution. Similarly, in the case of offloading the task to the EIN ($\mathcal{S}_{k,t} \in \mathcal{M}$ and $b_f^{(t)} = 1$), the total cost, $\mathcal{C}_{k,t} = \mathcal{C}_{k,f,t}^{EIN}$, entails the cost of transmission and execution. Thus, the joint optimization problem for the metaverse task computation and resource allocation can be formulated as follows:

$$\mathcal{J}_p : \min_{\mathcal{S}_{k,t}} \lim_{T \to \infty} \frac{1}{T} \sum_{t=1}^{T} \sum_{k \in \mathcal{K}} \mathcal{C}_{k,t} \quad (19)$$

$$\begin{aligned} s.t. \quad d1: & \sum_{v \in M \cup \{0\}} s_{k,v,t} \geq 1, \quad \forall k \in \mathcal{K}, v \in \mathcal{V}_k \\ d2: & \ b_f^{(t)} \in \{0,1\}, \\ d3: & \ (1 - b_f^{(t)}) T_{k,v,t}^{FIN} + b_f^{(t)} T_{k,v,t}^{EIN} = \mu, \\ d4: & \sum_{f \in} b_f^{(t)} V_f \leq \delta_{EIN}, \quad \forall t \in \mathcal{T}, \\ d5: & \sum_f (1 - b_f^{(t)}) V_f \leq \delta_{FIN}, \quad \forall t \in \mathcal{T}, \\ d6: & \ \mathcal{S}_{\ldots,t} \in \{0,1,\ldots,M\} \quad \forall k \in \mathcal{K}, \forall t \in \mathcal{T}, \\ & v \in \mathcal{V}_k. \end{aligned}$$

Constraint $d1$ states that at least one subtask of a given task is partially offloaded. Constraint $d2$ indicates that the offloaded task is executed at either the FIN or EIN. Constraint $d3$ enforces the task execution delay requirement. Constraints $d4$ and $d5$ indicate the EIN and FIN cache sizes, respectively. Constraint $d6$ represents the task computing modes where $\mathcal{S}_{k,t} = 0$ for the local task execution, and $\mathcal{S}_{k,t} = m (m \in \mathcal{M})$ indicates the TO to the EIN or FIN via channel $m$. Considering the dynamic user demand and network resource availability across different time slots coupled with the lack of the user transition request probability, it is tractable to solve problem $\mathcal{J}_p$ directly. Thus, we prove that the problem is NP-hard in Lemma 1.

*Lemma 1.* Problem $\mathcal{J}_p$, involving interactive TS from a user perspective and optimal TO for COIN across different time slots, is NP-hard.
*Proof.* See Appendix A.

### IV. DYNAMIC TASK OFFLOADING AND RESOURCE ALLOCATION

The optimization problem in Eq. (19) presents a unique challenge in determining an optimal solution in polynomial time, considering that dynamic system demand may arise as users engage in various metaverse activities. The primary difficulty is dealing with user offloading decisions based on the device constraints, the dynamic demands for the available offloading computing and cache resources, and the lack of certainty of the future demand while maintaining the metaverse QoE. To solve these challenges, we decomposed the problem into two parts to solve it: an optimal TSP on the user side and computational TOP on the COIN side. The TOP is modeled as a multiuser OPG, and a decentralized algorithm is proposed to address its NE. The MDP modeling is employed for the TSP, and the DDQN is used to learn the optimal policy for the TOP to meet unforeseen demand.

### A. MULTIUSER TASK-SPLITTING(TS) ALGORITHM

In any given time slot $t$, the TS decision is based on the cost of computing such a task and is unaffected by other slots. Each user's decision is independent and is unaffected by other users' decisions over time. Based on this, we formulate the problem as a noncooperative ordinal game and solve for its NE. Inspired by previous research [6], the game is modeled as follows.

We let $S_{k,t} = \{s_{k,0,t}, s_{k,1,t}, \ldots, s_{k,v,t} | s_{k,v,t} = m \ (m \in M)\}$ denote the set of TS strategies for user k where $v$ is the number of subtasks. The set of all UE strategies is given as $S_t = \{s_{k,t} | s_{k,t} \in S_{k,t}, k \in \mathcal{K}\}$, where $s_{k,t} = s_{k,0,t} = 0$ indicates the task is executed locally, and $s_{k,t} = s_{k,v,t} = m \ (m \in M)$ represents offloading to the EIN ($b_f^{(t)} = 0$) and FIN ($b_f^{(t)} = 0$) through channel $m$. Focusing on the UE perspective, the TSP and computational TOP for the multiuser game in slot t is given as

$$\mathcal{J}_{p1}: \min_{S_t} f_t(S_t) = \sum_{k \in \mathcal{K}} \mathcal{C}_{k,t}. \quad (20)$$
s.t. $d3$ and $d6$

Similar to previous work [6], where the TOP is considered, $S_t$ for the TSP has $(M+1)$ value selections, which is a combinatorial problem that is challenging to solve over a multidimensional discrete space $\{0, 1, 2, \ldots, M\}^{K \times v}$. Subsequently, we transformed it into a multiuser OPG and proved the existence of the NE to determine the optimal TS for each user.

Thus, the $\mathcal{J}_{p1}$ problem is transferred to a non-cooperative multiuser strategic game $g = \langle \mathcal{K}, \{S_{k,t}\}_{k \in \mathcal{K}}, f_t(\alpha_t) \rangle$, where the $\mathcal{K}$ is the game player set, the strategy of user $k$ is given as $S_{k,t}$, and $f_t(\alpha_t)$ denotes the computing cost for user $k$. The game objective is to achieve an NE solution $S_t^* = \{S_{1,t}^*, S_{2,t}^*, \ldots, S_{K,t}^*\}$, where no user can change the decision through cost reduction.

To split the task and determine the offloading location, a user in the game computes the cost of computing the subtask. The subtask is offloaded to the location with the least computing cost. In other words, a task is offloaded to either the FIN or EIN when its local computing cost is larger; $\mathcal{C}_{k,t}^L \geq (1 - b_f^{(t)}) \mathcal{C}_{k,t}^F + b_f^{(t)} \mathcal{C}_{k,t}^E$. By substituting Eqs. (5), (12), and (17) into the inequality

$$\delta_k^T T_k^L + \delta_k^E \mathcal{E}_k^L$$
$$\geq \frac{(I_f + V_f)(\delta_k^T + \delta_k^E \rho_k)}{\omega_{k,t}} + \delta_k^T T_{k,FIN}^{exe}$$
$$- b_f^{(t)}(\delta_k^T T_{k,FIN}^{exe} - \delta_k^T T_{k,EIN}^{exe})$$
$$\omega_{k,t} \leq \frac{(I_f + V_f)(\delta_k^T + \delta_k^E \rho_k)}{\delta_k^T (T_k^L - T_{k,FIN}^{exe}) + \delta_k^E \mathcal{E}_k^L + b_f^{(t)}(\delta_k^T T_{k,FIN}^{exe} - \delta_k^T T_{k,EIN}^{exe})},$$

the interference ($\mathcal{R}_{k,t}$) of user $k$ is derived as

$$\mathcal{R}_{k,t} = \sum_{n \in \mathcal{K}\{k\}, \alpha_{n,t} = \alpha_{k,t}} \rho_n \eta_n \leq$$
$$\frac{\rho_k \eta_k}{2^{B\left(\delta_k^T (T_k^L - T_{k,FIN}^{exe}) + \delta_k^E \mathcal{E}_k^L + b_f^{(t)}(\delta_k^T T_{k,FIN}^{exe} - \delta_k^T T_{k,EIN}^{exe})\right)} - 1} - \sigma^2. \quad (21)$$

The inference threshold ($\lambda_k$) for user $k$ is given as
$$\lambda_k = \frac{\rho_k \eta_k}{2^{B\left(\delta_k^T (T_k^L - T_{k,FIN}^{exe}) + \delta_k^E \mathcal{E}_k^L + b_f^{(t)}(\delta_k^T T_{k,FIN}^{exe} - \delta_k^T T_{k,EIN}^{exe})\right)} - 1} - \sigma^2. \quad (22)$$

Using Eq. (21), the user can reduce the energy consumption of the system—low interference suggests the user reduces the computation cost through offloading; otherwise, it is accomplished through local computing. Similar to previous work [43], the game $g$ is an OPG with a potential function:

$$\phi(S_t) = \frac{1}{2} \sum_k^K \sum_{n \neq k} \rho_k \eta_k \rho_n \eta_n \mathbb{1}(s_{n,t} = s_{k,t})$$
$$(s_{k,t} > 0) + \sum_{k=1}^K \rho_{k,} \eta_k \lambda_k \mathbb{1}(s_{k,t} = 0). \quad (23)$$

A user $k$ offloads a task when $\mathcal{R}_{k,t} \leq \lambda_k$, otherwise, it is accomplished by local computing. Fig. 3 illustrates the multiuser partial task splitting scenario in the network.

*Remark 1.* The TSP game $g$ is an OPG with a potential function and can achieve NE.
*Proof.* See Appendix B.

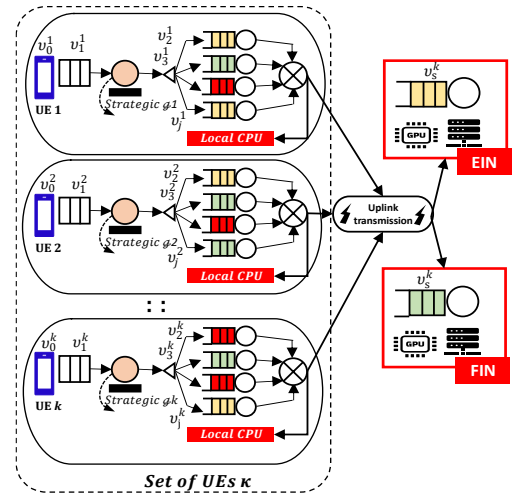

**FIGURE 3.** Multiuser partial task splitting using the game theory in the network.

Algorithm 1 solves the multi-user TSP from the user perspective by splitting the task into subtasks for computation offloading and achieving NE for the TS. The algorithms initialise the TS decision of all UE to 0 and execute a repeat-until loop until an END message is received. For each UE, the algorithm calculates the interference and transmission rate for each subtask $v \in \mathcal{V}_k$, then computes the corresponding optimal strategy for the TS by solving three constraints: $d1$, $d3$ and $d6$. These constraints enforce that the tasks are optimally decomposed to subtasks among the COIN resources while considering the energy consumption and delay constraints. Each user reaches an optimal $\mathcal{S}^*_{k,t}$ decision on its subtask $v \in \mathcal{V}_k$ by minimising the objective function $f_t(\mathcal{S}_{k,t}, \mathcal{S}_{-k,t})$. The users send requests to COIN for possible updates on the optimal offloading policy. The optimal policy is discussed in the subsequent section, addressing the TOP. Meanwhile, the COIN system sends the END message if no update request is sent. The users then offload their tasks after receiving the END message. Similar to [6], the convergence behaviour of Algorithm 1 is given in Lemma 2.

*Lemma 2.* The multi-user TS game $\mathcal{G}$ can reach an NE solution within a finite iteration given as $\frac{\frac{1}{2}K^2\Omega_{max}^2 + K(\Omega_{max}\lambda_{max} - \Omega_{min}\lambda_{min})}{\pi\Omega_{min}}|\pi \in \mathbb{R}^+$.

*Proof.* See Appendix C.

## B. DEEP REINFORCEMENT LEARNING-BASED TASK COMPUTATION OFFLOADING AND SOFTWARE CACHING

From the user's perspective, we established the TS strategies using OPG to determine a mutually satisfactory TS decision given a decomposable task in any time slot. By substituting the optimal $\mathcal{S}^*_t$ TS solution, the original problem can be transformed as follows:

$$\mathcal{J}_{p2} : \min_{\mathcal{S}_{k,t}} \lim_{T \to \infty} \frac{1}{T} \sum_{t=1}^{T} \sum_{k \in \mathcal{K}} \hat{\mathcal{C}}_{k,t} \quad (24)$$

s.t. $d1, d2, d4,$ and $d5$,

where

$$\hat{\mathcal{C}}_{k,t} = \sum_f \mathbb{1}(\mathcal{S}_{k,t} = 0) \mathcal{C}^L_{k,f,t}$$
$$+ \mathbb{1}(\mathcal{S}_{k,t} \in \mathcal{M})\left((1 - b_f^{(t)})\mathcal{C}^{FIN}_{k,f,t} + b_f^{(t)}\mathcal{C}^{EIN}_{k,f,t}\right). \quad (25)$$

The essence is to solve the optimal decision problem considering the demand from the users. For Q-learning, each subtask of every task for each user requires a state-action pair with the corresponding Q-values. With massive user demand, the memory cannot sustain such demands. To address the problem of providing an optimal offload policy in time slot $t + 1$, we applied the DDQN to capture the state-action pair and predict the Q-values corresponding to the optimal partial task offload based on time slot $t$. Problem $\mathcal{J}_{p2}$ is formulated as an MDP to design the DDQN algorithm. We elaborated on the state, action, and reward as follows.

- State: Due to the several users with several subtasks, the system state for each subtask can differ due to task input, computation, and software sizes; the availability of computing resources in FIN and EIN; and the decision of other users to offload partially, so the state space of a single agent increases exponentially with the number of pieces of UE. Therefore, to capture such heterogeneous states and solve the problem of state exploration, we propose the following equation to compute the system per user for all subtasks: $S_t = S_{t,K} \in (F + 1)^K$, where $S_{t,k \in K} = \sum_{f=0}^{F} S_f$, $S_f = S_{v \subseteq f} = \chi_0 S_0 + \chi_1 S_1 + \chi_2 S_2 + \chi_3 S_3 | \chi_{j \in \{0,1,2..\}} \in \{0,1\}$, and $S_{v \subseteq f}$ is the weighted sum of the user's CPU and cache, subtask input parameters $\langle I_f, V_f, P_f \rangle$, FIN/ EIN's CPU, and cache size. Using the proposed formula, we obtain a value reflecting each subtask's state.
- Action: The action in time slot $t$ is the offloading state in $A_t = b_{t+1} \in \{0,1,2\}^{F \times v}$.
- Reward: The reward is defined as the difference between the total current cost and total subsequent cost of the partial offloading policy of the system, given as $R_{t+1} = \sum_{k=1}^{K} \mathcal{C}_{k,t} - \sum_{k=1}^{K} \mathcal{C}_{k,t+1}$. The cost considers the queue at both the FIN and EIN and followsEqs. (10) and (15), respectively, for the delay cost calculation.

The objective of the problem is to minimize the total system cost in the network. The DDQN architecture employed in this study is depicted in Fig. 4. The model aims to capture and learn the users' state models and predict the optimal partial offloading policy, considering the total cost of the offloading policy in the system. To address the problem of high dimensionality and complex action space, we employed the state coding and action aggregation (SCAA) DNN model proposed in previous work [6], which is discussed below.

In the SCAA model, the influence of the input order on the DNN output is mitigated, where $X_t = \{\mathbb{1}(\mathcal{S}_{k,t} \in f) \, k \in K\}$ instead of the state $S_t$. The SCAA uses a dropout mechanism at the input layer for the code user's state and a two-layer output arch to aggregate the TO action dynamically.

Conventionally, the output of the neuron of the DNN corresponds to the number of all possible actions with each neuron and the output state-action value $Q(S_t, A_t)$. Considering the metaverse scenario and problem, assigning a neuron to each subtask from each user is impractical because this results in an exponential increase in the number of neurons with an increase in users. For example, $k$ users, with $F$ tasks decomposable into $v$ subtasks, requires $k \times F \times v$ neurons in the output layer.

In addition, listing all possible actions is impossible due to the heterogeneous size of the task input sizes and software. The two-layer architecture of the SCAA-DNN consists of $V$ neuron cells and one neuron for the last layer. The output layer $\Theta = (\Theta_1, \Theta_2 \dots \Theta_v)$ corresponds to the state-action value of computation and caching of the $v$ subtask. The last layer has no activation unit, but the output is the sum of all input variables. We let $w_L = (w_{1L}, w_{vL}, .. w_{VL})$ be the weight connection between the first and last layers. The state-action value of a specific action $A_t = b_{t+1}$ to $w_L$ is $w_{fL} = b_v^{(t+1)}, \forall v \subseteq f \in \mathcal{F}$. Therefore, the DNN outputs the predicted state-action value $Q(S_t, A_t) = \sum_{v \subseteq f \in \mathcal{F}} b_v^{(t+1)} \Theta$.

The DDQN is trained using the $\varepsilon$-greedy policy based on the NE DDQN, where the COIN agent explores the NE of the TS algorithm with probability $\varepsilon$ or determines the offloading strategies on $A_t^* = \arg\max_a Q(S_t, a)$ with a probability of $(1-\varepsilon)$. The NE DDQN explores NE and ensures the experiences in the memory pool are NE [33]. The users generate tasks decomposable into subtasks at time slot $(t+1)$ using Algorithm 1. Based on the subtasks from all users, the system state $S_t$, action $A_t$, reward $R_{t+1}$ and next state $S_{t+1}$ stored in the experience memory are used as training data for the DDQN. A batch of data in the experience memory are sampled in the form of $(S_t, A_t, R_{t+1}, S_{t+1})$.

During the training phase, $A_t$ assigned to $w_L$ is $w_{fL} = b_v^{(t+1)}, \forall v \subseteq f \in \mathcal{F}$, and the input is processed as $X_t = \{\mathbb{1}(v_j^k \in f): k \in K, j \in v\}$. The state-value action value is predicted and approximated as follows:

$$Q(S_t, A_t) = R_{t+1} + \gamma \max_a Q(S_{t+1}, a), \quad (26)$$

where $\gamma \in (0,1)$ is the discount factor. The target DNN is used to infer the value of $\max_a Q(S_{t+1}, a)$. The main network selects the action during the agent's interaction with the environment. The loss is calculated using the Huber function to stabilize the learning processing, whereas the DNN is trained using the backward algorithm. Algorithm 2 presents the training algorithm.

At the inference phase, we determined the optimal offloading policy in the time slot $(t+1)$ using the following optimal offloading statement:

$$\tilde{J}_{p2}: \max_{S_{k,t}} \sum_{v \subseteq f \in F} b_v^{(t+1)} \Theta_v \quad (27)$$

s.t. d7: $\sum_{v \subseteq f \in F} b_v^{(t+1)} V_v \leq C_i$

d8: $b_v^{(t+1)} \epsilon \{0,1,2\}$

d9: $i \in \{FIN, EIN, Local\}$.

The problem $\tilde{J}_{p2}$ is a typical knapsack problem. Unlike previous work [6], focusing on cache and noncache software in MEC, we derived the solution using the recursive algorithm that considers the cache availability in both the FIN and EIN. We let the matrix $\in (v, c)$ represent the optimal solution for the task using subtask $v \subseteq f \in \mathcal{F}$ using a cache size $c$ under a time constraint. The recursive function $\in (v, c)$ is defined as

$$\in (v, c) = \arg\min_{a \in \{0,1,2\}} \left(\in (v-1, c - b_v^{(t+1)} V_v) + b_v^{(t+1)} \Theta_v\right). \quad (28)$$

Algorithm 3 presents the optimal action state whose time complexity is $O(2FC + F)$.

The optimal offloading state in time slot $t+1$ assists the COIN agent in producing the optimal partial offloading at a low cost. Considering that the interference increases the overall cost when more subtasks are assigned to a specific computing resource, the FIN or EIN, the COIN agent learns to create an optimal partial offloading policy.

In summary, the algorithm takes the last layer out and initializes the variables of task location, weights, and total cost. Next, it iterates over the task software file and optimizes the knapsack, considering the file hit rate. The algorithm determines the optimal TO and caching policy based on the available capacity and previous caching decisions. Currently, the caching capacity, file size, and cache hit rates are used to determine an optimal partial offloading policy.

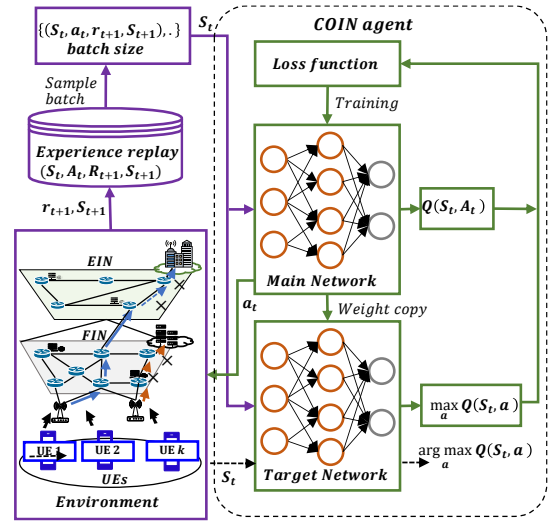

**FIGURE 4.** Task offloading optimization using the double deep Q-network (DDQN).

| **ALGORITHM 2**: *DDQN Training Algorithm* | |
|---|---|
| 1 | **Initialize:** Replay memory $RM$ to capacity $M$, the weight copy frequency $\mathsf{g}$ |
| 2 | **Initialize:** The main DNN with random weight $\theta$ and copy the weight $\theta$ to the target DNN |
| 3 | **For:** time slot $t = 1:T$ **do** |
| 4 | With probability $\varepsilon$, select an NE $A_t = S_t^*$ from the TS game; otherwise, select $A_t = \arg\max Q(S_t, a)$ as the partial offloading state in slot $t+1$. |
| 5 | Compute the reward $R_{t+1}$ using $A_t$ in time slot $t+1$. |
| 6 | Store transition $(S_t, A_t, R_{t+1}, S_{t+1})$ in the $RM$ |
| 7 | Sample random minibatch $(S_t, A_t, R_{t+1}, S_{t+1})$ from the $RM$ |
| 8 | Assign values to the weight $w_L$ based on the SCAA-DNN model |
| 9 | Assign $b_{t+1}$ to the weight of the TLA in the main net and obtain $Q(S_t A_t)$ |
| 1 | Using the loss function, perform gradient descent with respect to the DNN parameter |
| | Update the target network every $\mathsf{g}$ slots |
| 1 | **End for** |

**Algorithm 3.** Solving for an Optimal Action for Partial Offloading

**Input:** $\Theta_t, \{V_v \subseteq f, f \epsilon \mathcal{F}\}$

**Output:** The optimal partial offloading policy $\boldsymbol{b}_{t+1}$

1    $\boldsymbol{b}_{t+1} = [0]_F, \in = [0]_{F \times C}, \in_r = [0]_{F \times C};$
2    **for** each $v \epsilon [1, F]$: **do**
3      **if** $v < F$ **then**
4        **for** each $c \epsilon [1, C]$: **do**
5          **if** $v == 1$ **then**
6            $\in_r (v, c) = \mathbb{1}(V_v < c)$
7            $\in (v, c) = \in_r (v, c)\Theta_v$
8          **else**
9            $\in_r (v, c) = \arg\min_{a \in \{0,1,2\}} (\in (v - 1, c - aV_v) + a\Theta_v)$
10            $\in (v, c) = \in_r (v, c)\Theta_v + \in (v - 1, c - \in_r (v, c)V_v)$
11          **end if**
12        **end for**
13      **else**
14        $\in_r (F, C) = \arg\min_{a \in \{0,1,2\}} (\in (F - 1, C - aV_v) + a\Theta_F)$
15        $\in (F, C) = \in_r (F, C)\Theta_F + \in (F - 1, C - \in_r (F, C)V_F)$
16      **end if**
17    **end for**
18    $\boldsymbol{b}_{t+1}(F) = \in_r (F, C)$
19    **for** each $v = F - 1: -1: 1$ **do**
20      $\boldsymbol{b}_{t+1}(v) = \in_r (v, C - \sum_{v+1 \le j \le F} \boldsymbol{b}_{t+1}(j) * V_i)$
21    **end for**
     **return** $\boldsymbol{b}_{t+1}$

## V. SIMULATION RESULTS

The This section presents extensive simulations to verify the effectiveness of the proposed TS using OPG and the subsequent optimization of the TS solution using the DDQN over time, based on the dynamic system state, to improve overall performance. We considered a scenario with $K = 30$ users randomly distributed within a $200 \times 200$ m cell region, and the FIN and EIN were placed at the center and edge of the cells, respectively. The FIN is closer to the user and equidistant to the EIN, which is placed after the FIN.

As presented in Fig. 1, we assume the number of subtasks is $J = \{0,1,2,3\}$ where 0, 1, 2, and 3 represent object tracking and detection, SLAM with the point cloud dataset, hand gesture and pose estimation, and multimedia processing and transport, respectively. The input parameter data size $I_f$ for each task $f$ for user $k$ is uniformly randomly selected for each subtask $f_k = \{f_j: j \in J\}$ in $[1, I_{jmax}]$ GB. Similarly, the software data size and CPU cycle for computing the subtask are uniform and randomly selected in $[1, V_{jmax}]$ GB and $[1, P_{jmax}]$ gigacycles.

We assumed each metaverse task $f \epsilon \mathcal{F}$ is denoted by the tuple parameters: $\langle I_f, V_f, P_f \rangle$, where $I_f, V_f,$ and $P_f$ denote the task input size, software volume, and computational load, respectively. The task $f = \{f_{k,0}, f_{k,1}, f_{k,2} \ldots f_{k,j}\}$, where $k$ is the user and $j$ is the number of subtasks. Similar to previous work [6, 43], the energy coefficient is $5 \times 10^{-27}$. Table 3 presents the simulation environment settings unless otherwise stated.

For the evaluation, we considered the baselines for which RL is not used to optimize the TS. The following baseline strategies were considered for the evaluation.

1) **MEC:** In this strategy, subtasks are computed locally or offloaded to MEC based on the available caching memory. This baseline enables observing the performance of the COIN based on the proposed approach against the MEC baseline. This strategy is among the state-of-the-art strategies. For instance, one study [23] suggested this baseline through a greedy strategy to reach an offloading decision by choosing the execution location with minimal overhead.

TABLE II
PARAMETER SUMMARY

| Parameter | Value |
| --- | --- |
| User number: $K$ | 30 |
| Subtask number: $\mathcal{V}_k$ | 10 |
| Time slot: $T$ | 1,000 |
| Wireless transmission bandwidth: $B$ | 50 MHz |
| Transmission power: $\rho_k^{EIN}/\rho_k^{FIN}$ | 0.5 W |
| Guassain noise variance: $\sigma^2$ | $2 \times 10^{-13}$ |
| CPU capability of user $k$: $F_k^L$ | 1 GHz |
| CPU capability of FIN: $F_k^{FIN}$ | 60 GHz |
| CPU capability of EIN: $F_k^{EIN}$ | 100 GHz |
| Cache size of FIN: $\delta_{FIN}$ | 3 GB |
| Cache size of EIN: $\delta_{EIN}$ | 5 GB |
| Number of channels: $M$ | 10 |
| DNN learning rate | 0.0008 |
| Experience replay memory size: $E$ | 10,000 |
| Batch size | 32 |
| Discount factor: $\gamma$ | 0.9 |
| $I_{max}, V_{max}, P_{max}$ | 5–10 |
| $r_{min}, r_{max}$ | 1, 10 |

2) **Random Computing:** This strategy is based on random partial offloading of subtasks to the FIN and EIN or locally computed on the device. Several studies have employed this strategy to demonstrate the efficiency of offloading algorithms [45]

4) **OPG:** This strategy considers a scenario where RL does not optimize the TS solution. This baseline grants insight into the overall network cost when the DDQN is not applied. Some state-of-the-art methods have proposed this approach [36, 43].

We employed the OPG specifically for TS; thus, each user conducts the game among its subtasks using the interference generated by each subtask to reach a decision.

To compare the proposed approach's performance fairly, we analyzed several aspects. We compared average costs and rewards across training episodes against benchmarks. Fig. 5(a) illustrates that our model consistently yielded the lowest costs after 200 episodes, with cost savings surpassing 13% over the MEC approach by the 1000th episode. Fig. 5(b) shows that the proposed method achieved a higher system reward, surpassing MEC by about 16%. This outcome indicates that the proposed method can offer an optimal offloading strategy under changing network conditions.

Furthermore, we assessed the proposed method's efficiency with user numbers ranging from 20 to 45. The analysis, illustrated in Fig. 6, shows that the proposed method outperforms others in cost efficiency across different user numbers, achieving up to a 23% cost reduction compared to the MEC, next best method, when user numbers increase to 45. As user numbers grow, the proposed method consistently decreases system costs, highlighting its robust scalability and promising application for expansive user networks. Thus, our method is effective for both small-scale and large-scale user scenarios.

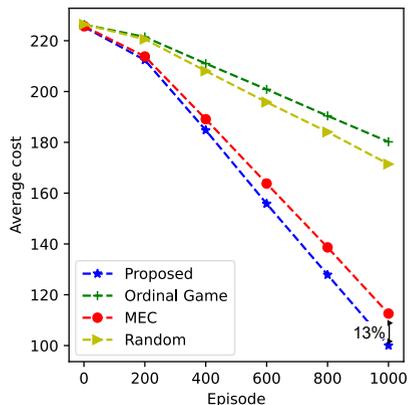

(a)

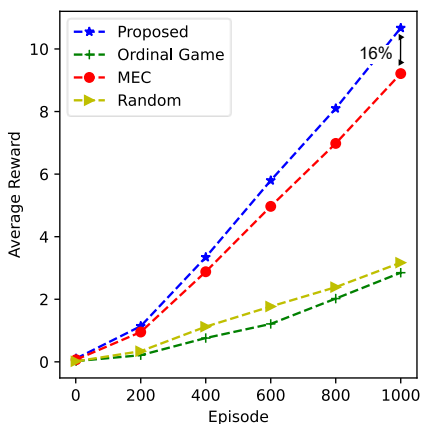

(b)

**FIGURE 5.** Performance evaluation with respect to the iteration step under various episodes: (a) system cost and (b) system reward.

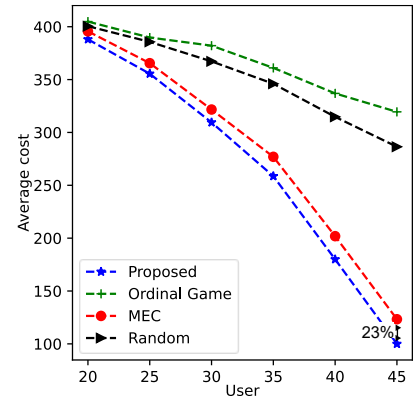

**FIGURE 6.** Comparison of the average cost for each time slot for different numbers of users.

In assessing the system model's performance for compute-intensive tasks, we varied the values of $V_{max}$ and $P_{max}$ to analyze their influence on cost efficiency. At a $V_{max}$ value of 26, our model achieves a cost reduction of 24% compared to the second-best model, the MEC, as shown in Fig. 7. Moreover, the analysis of $P_{max}$, as depicted in Fig. 8, indicates that our model attains a maximum cost efficiency improvement of 45% at a $P_{max}$ of 14. These outcomes consistently demonstrate the robustness of our model in maintaining superior cost-efficiency gains across varied task input. It highlights its potential for adaptability and reliable performance in scenarios with diverse computational workload demands.

Finally, we consider evaluating the proposed approach's performance under four, seven, and 10 subtasks. The user number and training episodes were 30 and 2,000, respectively. Fig. 9 illustrates the optimal subtask computational offloading distribution and the corresponding cost for the different subtasks. As illustrated in the distribution, the OPG method offloads most of the subtasks from the users to the FIN or EIN, suggesting that each player in the game independently determines its TS irrespective of the general demand by other users. Similarly, random offloading distributed tasks marginally between the user device and the FIN and EIN. However, the OPG and random methods incur a high average cost on the overall system.

Furthermore, in the MEC approach, subtasks are partially offloaded to the EIN only at the network edge. Although this approach has a significant cost reduction compared to the OPG and random methods, performing some computations at the FIN in the proposed approach further reduces the cost. This approach further reduces the cost by 8%, 5%, and 70% for four, seven, and 10 subtasks, respectively. This outcome reveals that this approach is more suitable for multiple subtask offloading, particularly for metaverse scenarios where multiple subtasks are expected.

### A. Discussion

In the scenario by Ding, et al. [23], UE has multiple MEC servers to choose from in a one-to-many fashion. The UE must determine the MEC servers within the communication range. Once it is determined that the servers have available computing resources, the UE continues offloading to that location. However, this method considers the dynamic user demands and changing resource availability of other servers once offloading begins. The handling or optimization of the offloading policy is necessary to balance the network load considering the datelines of the tasks in the applications, such as the metaverse. This approach addresses this problem by constantly updating the offloading policy to handle unforeseen demand. Additionally, UE acts with a greedy strategy; therefore, the optimal offloading strategy must be updated for better performance over time. The proposed approach reduces the overall cost by 70% and improves the reward by 8%.

Unlike previous work [36, 43] that employed the game theory for multiuser computational offloading, where each player's decision is relative to the others considering the inference, the proposed game for each player is centered on the subtasks and interference generated as a result. As suggested by the results, this approach is suitable for multiple subtask situations because it reduces the overhead cost by 70% for 10 subtask scenarios. Thus, the proposed DDQN for updating the offloading policies addresses the dynamic user demand and network resources status.

## VI. CONCLUSION

This paper investigated partial TO and caching problems for metaverse divisible tasks in a dynamic multiuser COIN to minimize costs based on energy and latency. We solved the problem by decomposing it into two parts—the first deals with the TSP from the user perspective, and the second deals with the optimal TO from the network perspective. We performed extensive simulations to verify the effectiveness of the proposed TS solution using OPG and the TO solution using the DDQN. We reformulated the TSP as a multiuser game and proposed a decentralized algorithm to solve for the NE solution. Next, we proposed a DDQN to optimize the TS solution over time, considering the changing system state to improve the overall performance. Unlike the conventional DDQN algorithm, where intelligent agents sample offloading decisions randomly and from the main network s within a probability, in the proposed DDQN algorithm, the COIN intelligent agent uses offloading where the COIN agent explores the NE of the TS algorithm and the DNN within the probability.

By exploring the TS game solution, the agent can update its policies to make more informed decisions in the future, leading to improved performance over time. Furthermore, we formulated a system state that captures the heterogeneous nature of subtasks for UE, FIN, and EIN computational resources and communication conditions. The proposed approach performs better in lowering partial TO costs and optimizing future offloading policies against baseline approaches. As COIN is likely to coexist with existing edge computing, future studies can explore partial subtask offloading in a collaborative scenario where the task type, dependency, and software caching status can be investigated for optimal partial offloading strategies.

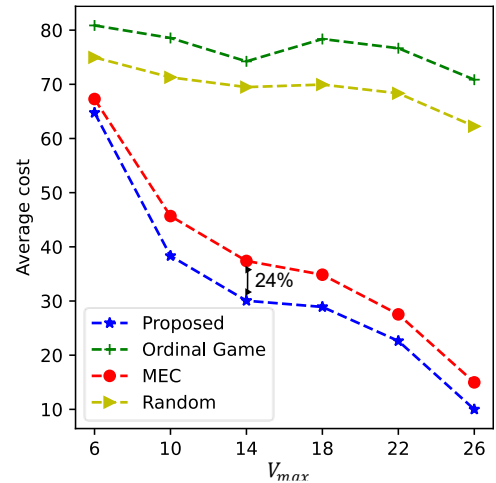

**FIGURE 7.** Comparison of the average cost for each time slot for different maximum software volume $V_{max}$.

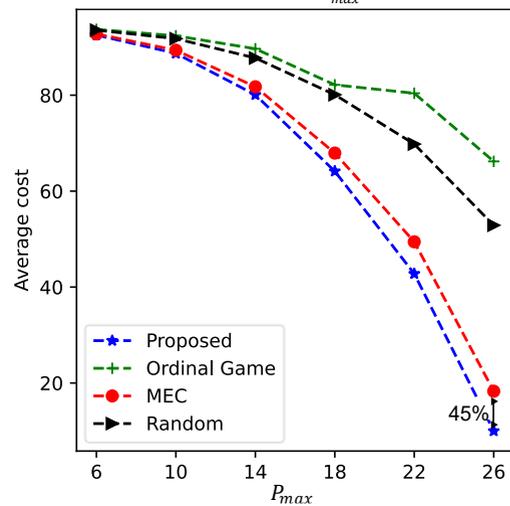

**FIGURE 8.** Comparison of the average reward for each time slot for different maximum computation load $P_{max}$.

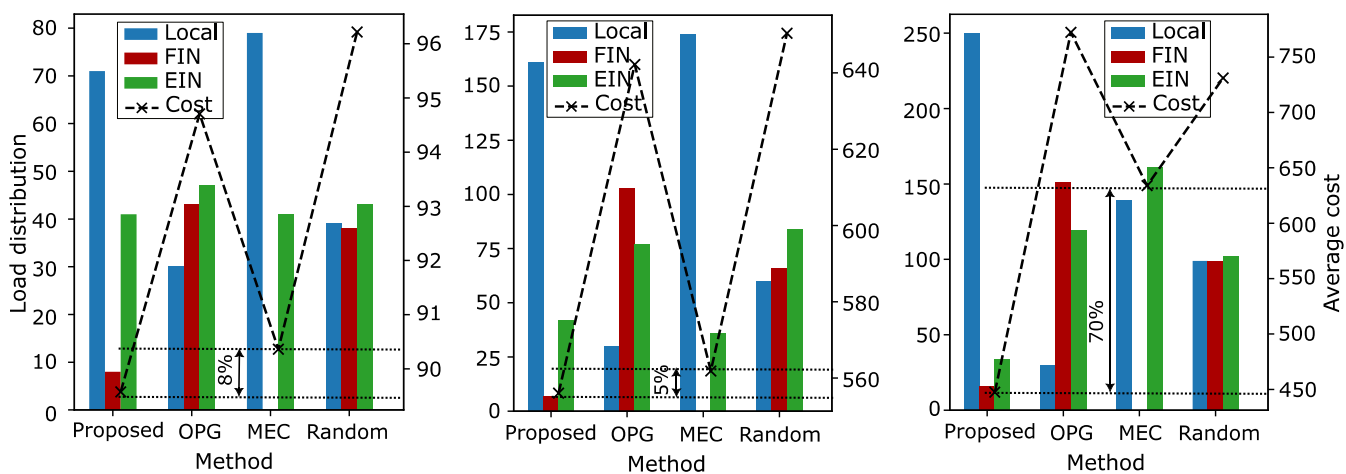

**FIGURE 9.** Comparison of optimal partial computational offloading distributions under various numbers of subtasks: (a) Four subtasks, (b) seven subtasks, and (c) 10 subtasks.

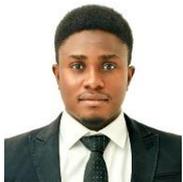

**IBRAHIM ALIYU** (Member, IEEE) received a PhD in computer science and engineering from Chonnam National University in Gwangju, South Korea, in 2022. He also holds BEng (2014) and MEng (2018) degrees in computer engineering from the Federal University of Technology in Minna, Nigeria. He is currently a postdoctoral researcher with the Hyper Intelligence Media Network Platform Lab in the Department of ICT Convergence System Engineering at the Chonnam National University. His research focuses on distributed computing for massive metaverse deployment. His other research interests include federated learning, data privacy, network security, and artificial intelligence for autonomous networks.

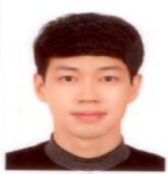

**Seungmin Oh** received a bachelor's degree in the Department of Digital Contents from the Korea Nazarene University in 2019. He received his M.S. degree in the Department of ICT Convergence System Engineering from the Chonnam National University in 2021. He has been pursuing his Ph.D. in the Department of ICT Convergence System Engineering at Chonnam National University since 2021. His current research interests include deep learning and machine learning, and computer vision.

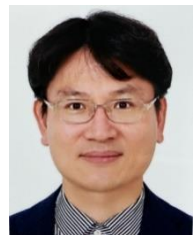

**NAMSEOK KO** is currently a director leading the Mobile Core Network Research Section at the Electronics and Telecommunication Research Institute (ETRI). He is also an associate professor with the Department of Information and Communication Engineering at the University of Science and Technology (UST). He received MS and PhD degrees from the Korea Advanced Institute of Science and Technology (KAIST) in Daejeon, South Korea, in 2000 and 2015, respectively. He previously served as a vice chair with the ITU-T Focus Group IMT-2020. He is currently a vice chair with SG11 and the Technology Committee and a rapporteur for Q.20 of SG13 in ITU-T and leads the Network Technology Working Group at the 6G Forum in Korea. Since joining the ETRI in 2000, he has participated in various R&D projects, including 5G mobile core network technology. He is currently leading several ongoing projects related to 6G network architecture.

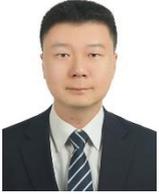

**TAI-WON UM** received a BS degree in electronic and electrical engineering from Hongik University in Seoul, South Korea, in 1999 and MS and PhD degrees from the Korea Advanced Institute of Science and Technology (KAIST) in Daejeon, South Korea, in 2000 and 2006, respectively. From 2006 to 2017, he was a principal researcher with the Electronics and Telecommunications Research Institute (ETRI), a leading government institute on information and communications technology in South Korea. He is currently an associate professor at Chonnam National University in Gwangju, Korea. He has been actively participating in standardization meetings, including ITU-T SG20 (Internet of Things, smart cities, and communities).

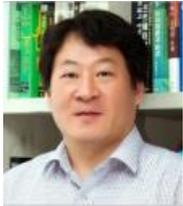

**Jinsul Kim** received a BS degree in computer science from the University of Utah at Salt Lake City in Utah, USA, in 1998 and MS (2004) and PhD (2008) degrees in digital media engineering from the Korea Advanced Institute of Science and Technology (KAIST) in Daejeon, South Korea. Previously, he worked as a researcher at the Broadcasting/Telecommunications Convergence Research Division, Electronics and Telecommunications Research Institute (ETRI) in Daejeon, Korea, from 2004 to 2009 and was a professor at Korea Nazarene University in Cheonan, Korea, from 2009 to 2011. He is a professor at Chonnam National University, Gwangju, Korea. He is a member of the Korean national delegation for ITU-T SG13 international standardization. He has participated in various national research projects and domestic and international standardization activities. He is a co-research director of the Artificial Intelligence Innovation Hub Research and Development Project hosted by Korea University and is the director of the G5-AICT research center.